\documentclass[twocolumn]{svjour3}         
\smartqed  
\usepackage{graphicx}
\usepackage{mdwlist}
 \usepackage{mathptmx}      
 \usepackage{algorithm2e}
\begin{document}

\title{Collective Classification of Textual Documents by Guided Self-Organization in T-Cell Cross-regulation Dynamics}


\author{Alaa Abi-Haidar \and Luis M. Rocha\\
}


\institute{Alaa Abi-Haidar \and Luis M. Rocha  \at
School of Informatics and Computing, Indiana University, Bloomington IN 47401, USA\\
FLAD Computational Biology Collaboratorium, Instituto Gulbenkian de Ci\^{e}ncia, Portugal \\
              \email{aabihaid@indiana.edu ; rocha@indiana.edu} }           

\date{Received: date / Accepted: date}

\maketitle

\begin{abstract}



We present and study an agent-based model of T-Cell cross-regulation in the adaptive immune system, which we apply to binary classification.
Our method expands an existing analytical model of T-cell cross-regulation \cite{carneiro2007tnc} that was used to study the self-organizing dynamics of a single population of T-Cells in interaction with an idealized antigen presenting cell capable of presenting a single antigen.
With agent-based modeling we are able to study the self-organizing dynamics of multiple populations of distinct T-cells which interact via antigen presenting cells that present hundreds of distinct antigens.
Moreover, we show that such self-organizing dynamics can be guided to produce an effective binary classification of antigens, which is competitive with existing machine learning methods when applied to biomedical text classification.
%

%

More specifically, here we test our model on a dataset of publicly available full-text biomedical articles provided by the BioCreative challenge \cite{bc2.5}.
We study the robustness of our model's parameter configurations, and show that it leads to encouraging results comparable to state-of-the-art classifiers.
Our results help us understand both T-cell cross-regulation as a general principle of guided self-organization, as well as its applicability to document classification. Therefore, we show that our bio-inspired algorithm is a promising novel method for biomedical article classification and for binary document classification in general.

\end{abstract}

\section{Background}
\label{intro}
At least since the beginning of systematic genomic studies, there has been a
tremendous growth of scientific publications in the life sciences
\cite{hunter2006blp}. Pubmed (\texttt{\small{http://pubmed.gov}}) now contains a growing collection
of more than 19 million biomedical articles.
Manually classifying these articles as relevant or irrelevant to a given topic
of interest  is very time consuming and inefficient for curation of new
published articles \cite{Jensen2006}.
Literature (or text) mining offers solutions for automatic biomedical document
classification and information extraction from huge collections of text, as
well as the linking of numerous biomedical databases and knowledge resources
\cite{Jensen2006,Shatkay03JCB}.
Because it is very important to validate and assess the quality of proposed
solutions, various community-wide competitions and challenges have been
organized so that automatic systems can be evaluated against human annotated
data sets (e.g. TREC Genomics \cite{bibliome_hersh04}).
One such effort is the BioCreative challenge, which aims to assess
biomedical literature mining in real-world scenarios
\cite{biocreative_issue,Krallinger_BC2_IAS,bc2.5}. Machine learning has
offered a plethora of solutions to this problem
\cite{Jensen2006,feldman2006tmh}, however, even the most sophisticated of
solutions often overfit to the training data and do not perform as well on
real-world scenarios such as that provided by BioCreative
\cite{abihaidar_GB08,casci_TCBB_10}.
One of the challenges of biomedical article classification in real-world
scenarios is the presence of highly unbalanced classes; typically, there are
many more irrelevant than relevant documents, without prior knowledge of class
proportions.
This was the case of the article classification data set in the Biocreative
BC2.5 challenge \cite{bc2.5}.
While participating teams (including our own team \cite{casci_TCBB_10}) did
not enter bio-inspired solutions, the unbalanced nature of classes and the
presence of conceptual drift, which we showed to occur between training to
test data sets \cite{abihaidar_GB08,casci_TCBB_10}, may be a good scenario
to test classifiers inspired by the vertebrate immune system---which must
operate under class-imbalance with permanent drift in the populations of
pathogens encountered.
Therefore, here we explore the feasibility of using T-Cell cross-regulation
dynamics to classify biomedical articles using the real-world scenario provided by
the Biocreative 2.5. data set.

The immune system (IS) is a complex biological system made of millions of cells all
interacting to distinguish between
self and nonself 
substances,
%
%
to ultimately attack the latter \cite{hofmeyr2001iii}\footnote{We use the
terminology of self/nonself discrimination, though perhaps a more accurate
description is classification of harmless vs. harmful substances; harmless can also include antigens from bacteria that are necessary for
vertebrate bodies, and harmful can also include body's own tumor cells.}.
In analogy, relevant
biomedical articles for a given concept need to be distinguished from
irrelevant ones.
To perform such a topical
classification, we can use the occurrence and co-occurrence of thousands of
words in a document.
In this sense, words can be seen as interacting in a text in such a way as to
allow us to distinguish between relevant and irrelevant documents---in analogy
with the interactions among T-cells and antigens that lead to self/nonself
discrimination in the immune system, as we describe below.

Our approach is based on the idea that the immune system is a distributed collection of molecular constituents with no central controller \cite{Segel&Cohen01}. Therefore, immune classification needs to result from a \emph{collective classification} process, defined as the
ability of decentralized systems of many components to classify situations that require global
information or coordinated action \cite{Mitchell:2006fk}.
Nature is full of examples of collective classification: the dynamics of stomata cells on leaf surfaces are known to be statistically indistinguishable from the dynamics of automata that are capable of performing nontrivial classification \cite{Peak:2004fk}, biochemical intracellular signal transduction networks are capable of emergent classification
\cite{Helikar:2008fk}, quorum sensing in bacteria \cite{Quorum_Sensing_Walters2006} and social insects \cite{quorum_sensing_ant_Pratt2005}, etc.
We can also study collective classification in general models of complex systems such as Cellular Automata, namely by identifying regular patterns in the dynamics that store, transmit and process information \cite{Crutchfield&Mitchell95,Rocha2005a,Shalizi:2006fk}.
Here, instead of looking at general models of complex systems, we focus on a
specific immunological model of T-Cell cross-regulation dynamics
\cite{carneiro2007tnc}. We are are interested in exploring the collective
dynamics of this model to: (1) build a novel bio-inspired machine learning
solution for document classification, and (2) understand how well collections
of T-Cells engaged in cross-regulation perform as a classifier. The first goal
entails a bio-inspired approach to computational intelligence, and the second a
computational biology experiment, but both are based on artificial life
principles.

It should be noticed that recent work in artificial immune systems (AIS)
\cite{timmis2007ais} has lead to a few immune-inspired solutions to document
classification in general \cite{twycross2002isa}, however, none to our
knowledge has been applied to biomedical article classification nor does any employ T-cell cross-regulation dynamics.
There are several reasons why T-Cell cross-regulation is appealing to explore for classification tasks.
Dasgupta and Nino \cite{dasgupta2008ict} concluded that negative selection algorithms suffer from scalability (for binary representation) and dimensionality issues (for real-valued representation), while algorithms inspired by clonal selection and artificial immune networks have been shown to be equivalent or very similar to evolutionary algorithms, with antibody somatic hypermutation instead of genetic variation \cite{garrett2003pne}. As we show below, our novel model for text classification, in addition to promise in imbalanced and dynamic scenarios, is scalable and capable of dealing with large numbers of textual features.

We have already proposed an agent-based model of T-cell cross-regulation for
spam detection \cite{abihaidar_icaris_08,abihaidar_alifexi_08}. Our
distributed model extends the original analytical model of T-Cell
cross-regulation dynamics \cite{carneiro2007tnc} to be able to deal with many
multiple features simultaneously, and therefore render the model applicable to
real-world applications.
Our results on spam-detection were comparable to state-of-art text classifiers
\cite{abihaidar_icaris_08,abihaidar_alifexi_08}.
However, our initial agent-based implementation of cross-regulation dynamics
did not explore important parameter configurations such as the death rate of
T-cells or the best training strategies. It also lacked an extensive parameter
search for optimized performance. Here, we address some of these issues on full-text biomedical data from BioCreative \cite{bc2.5}.

First, we study the  effect of cell death on the dynamics of T-cell cross-regulation and its importance for improving classification performance.
We also study the effect of training exclusively on relevant or positive documents.
This is relevant to understand immune classification dynamics, because in the process of T-Cell maturation, to prevent autoimmunity, T-Cells are checked exclusively against self epitopes---eliminating T-Cells that bind to self.
In the context of machine learning, this is similar to what is known as positive unlabeled (PU) training, which we test here against training on both relevant (positive) and irrelevant (negative) documents.
Next, we study the importance of the original temporal sequence of bio-medical articles. Text mining classifiers do not typically depend on the sequence of documents they are trained with, but our model of T-cell cross-regulation dynamics does. Therefore, we are interested in ascertaining if the sequence-dependence of ensuing collective dynamics can be used to track the natural change in real-world textual corpora, i.e. concept drift \cite{Tsymbal2004}.
Finally, we also study the effect of biases in the initial T-cell population. This more extensive study allows us to better understand the behavior of T-cell cross-regulation dynamics and  establish its capability to classify sequential data. It also leads to a competitive, novel bio-inspired text classification algorithm.

In the next section we give an introduction to the vertebrate immune system. In section \ref{CRM}, we discuss the existing analytical model of  T-cell cross-regulation. In section \ref{ABCRM}, we present our agent-based model of T-Cell cross-regulation for binary classification, here applied to document classification. In section \ref{datas}, we describe the biomedical data provided by Biocreative and the feature selection method. In section \ref{results}, we study the robustness of our model on various parameter ranges and experimental setups. Finally, in section \ref{valid}, we compare our model with state-of-art classifiers.

\section{The Immune System as Inspiration}
\label{IS} The vertebrate adaptive immune system\footnote{A good, though already a
bit dated, overview of the vertebrate immune system for the artificial life
community is Hofmeyer's \cite{hofmeyr2001iii}.} (IS) is a complex network of
cells that distinguishes between self and nonself substances or
antigens---usually fragments of proteins that can be recognized by the immune
system.
When nonself antigens are discovered, an immune response to eliminate them is
set in motion. Recognizing self antigens, which obviously should not lead to an
(auto)immune response to eliminate them, is resolved by negative selection of
T-cells which takes place in the thymus, and removes T-Cells that strongly bind
to self antigens---after positive selection of T-Cells that are capable of
binding with the major histocompatibility complex (MHC) \cite{paul1993fi}.
It is in the thymus that T-cells develop and mature; only T-cells that have
failed to bind to self antigens are released (as mature naive T-cells), while the rest
of the T-cells is culled. Mature T-cells are allowed out of the thymus to
detect nonself antigens. They do this by binding to antigen presenting cells
(typically B-cells, macrophages and dendritic cells) that collect and present
antigens via MHC after breaking them by lysosome. The specific T-cells that are
able to bind to the presented antigens then stimulate B-cells that start a
cascade of events leading to antibody production and the destruction of the
pathogens or tumors linked to the antigens.
However, it is possible that T-cells and B-cells, which are also trained in the
thymus and bone marrow, mature before being exposed to all self antigens. Even
more problematic is the somatic hypermutation that ensues in lymph nodes after
the activation of B-cells through a process known as ``clonal selection'' \cite{burnet1959clonal}. At this stage, it is possible to generate many
mutated B-Cell clones that could bind to self antigens. Either situation can
cause auto-immunity by generating T-cells capable of attacking self antigens.
One way to deal with this problem is by a process called costimulation which
involves the co-verification of self antigens by both T-cells and B-cells
before the antigen is identified as associated with a nonself pathogen to be
attacked. To further insure that the T-cells do not attack self, another type
of T-cells known as regulatory T-cells, are formed in the thymus where they
mature to avoid recognizing self antigens. These regulatory T-cells have the
responsibility of preventing autoimmunity by down-regulating other T-cells that
might bind and kill self antigens.
Our model is based on this process of T-Cell cross-regulation.

\begin{figure}
\begin{center}
\includegraphics[width=9cm,height=8cm]{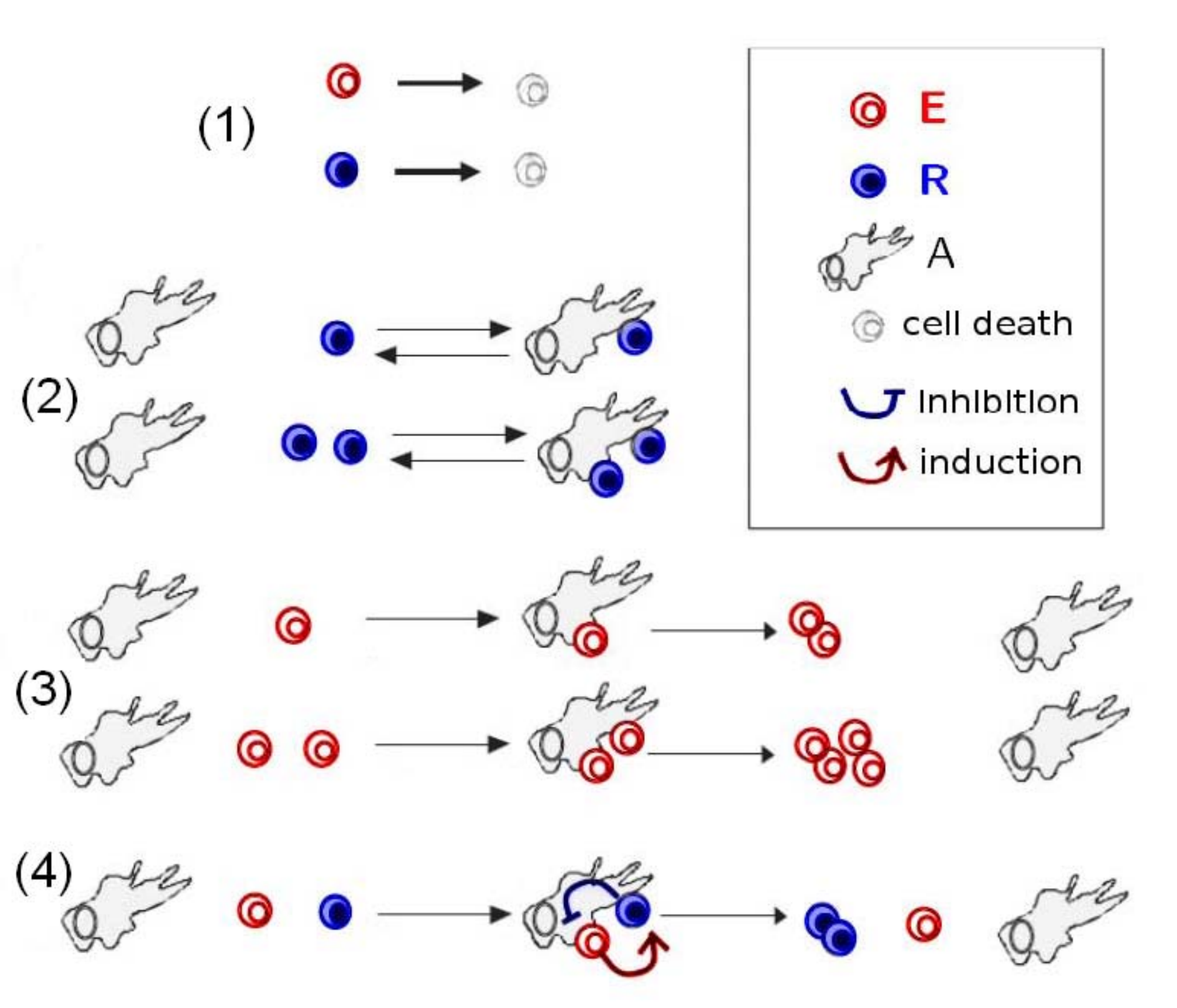}
\end{center}
\caption{ CRM interactions that define the dynamics of APC and $E$ and $R$ T-cells. The model assumes that  APC can only form conjugates with a maximum of two T-cells. Adapted from \cite{carneiro2007tnc}. }
\label{fig:crm}
\end{figure}

Artificial Immune Systems (AIS) are artificial life tools, inspired by theories
and components of the immune system, and applied towards solving computational
problems, such as categorization, optimization and decision making
\cite{decastro}. Common AIS techniques are based on specific theoretical
models explaining the behavior of the IS such as: Negative Selection, Clonal
Selection, Immune Networks and Dendritic Cells \cite{timmis2007ais}. AIS fall
in categories: (1) mathematical and computational models to understand IS
behavior and (2) engineering of adaptive machine learning algorithms. While our
approach fits more immediately under the second category, our goal is also to use
our classifier to test the prevailing model of T-cell cross-regulation and
therefore also contribute to the first category of the study of AIS.

\section{The Cross-Regulation Model}
\label{CRM} The \emph{T-cell Cross-Regulation Model} (CRM)
\cite{carneiro2007tnc} is a dynamical system that aims  to distinguish between
self and nonself protein  fragments  (antigens)  using only four possible
interaction rules amongst three cell-types: \emph{Effector T-cells} ($E$),
\emph{Regulatory T-cells} ($R$) and \emph{Antigen Presenting Cells} (APC). As
their name suggests, APC present antigens for the other two cell-types, $E$ and
$R$, to recognize and bind to them. Effector T-cells ($E$) proliferate upon
binding to APC, unless adjacent to regulatory T-cells ($R$), which regulate $E$
by inhibiting their proliferation. For simplicity, proliferation of cells is
limited to duplication in quantity in contrast to having a proliferation rate.
T-cells that do not bind to APC die off with a certain death rate. The dynamics
of the CRM depend on four interaction rules defined by the following reactions
(illustrated in Fig. \ref{fig:crm}):

\noindent
\begin{eqnarray}
&& E_{\overrightarrow{d_E}} \{\} \mbox{ and } R_{\overrightarrow{d_R}} \{\} \\
&& A+R \rightarrow A+ R \\
&& A+E \rightarrow A+ 2E \\
&& A + E + R \rightarrow A+ E + 2R
\label{4l}
\end{eqnarray}

\noindent Reaction (1) defines $E$ and $R$ apoptosis with the corresponding
death rates $d_E$ and $d_R$. The last three proliferation reactions define the
maintenance of $R$ (2), the duplication of $E$ (3), and the maintenance of $E$
and duplication of $R$ (4).

Carneiro et al \cite{carneiro2007tnc} developed  the analytical CRM to study the dynamics of a single population of T-cells (with effector and regulatory elements) that interacts with APC that present a single antigen. In \cite{abihaidar_icaris_08,abihaidar_alifexi_08}, we extended the original CRM model to be able to deal with multiple populations of antigens and T-Cells using agent-based modeling.
%
%
More recently,  Sepulveda \cite[pp 111-113]{nuno} extended the original CRM to study analytically multiple populations of T-cells that recognize antigens presented by APC capable of presenting at most two distinct antigens.
In our model, explained in detail in the next section, APC are capable of presenting hundreds of antigens to be recognized by T-cells of hundreds of different populations, using the same four interaction rules of the CRM.

\section{The Agent-Based Cross-Regulation Model}
\label{ABCRM}

In order to adapt the CRM to an \emph{Agent-Based Cross-Regulation Model} (ABCRM)
for text classification, one has to think of documents as analogous to the
organic substances that upon entering the body are broken into constituent
fragments. These fragments, known as epitopes, are presented on the surface of
Antigen Presenting Cells (APC) as antigens. In the current application of the ABCRM, antigens are  textual
features (e.g. words, bigrams, titles, numbers) extracted from articles and
presented by artificial APC such that they can be recognized by a number of
artificial Effector T-cells ($E$) and artificial Regulatory T-cells ($R$).
Individual $E$ and $R$ have receptors for a single, specific (textual) feature:
they are \emph{monospecific}. $E$ proliferate\footnote{The simplification of proliferation to mere duplication adopted in the canonical CRM model is maintained in our agent-based model to minimize the number of parameters (excluding proliferation rates) and the parameter search space} upon binding to antigens
presented by APC unless suppressed by $R$; $R$ suppress $E$ when binding in
adjacent locations on APC. Individual APC present various document features:
they are \emph{polyspecific}. Each APC is produced when documents enter the
artificial cellular dynamics, by breaking the former into constituent textual
features. Therefore we can say that APC are representative of specific
documents whereas $E$ and $R$ are representative of specific features.

\begin{figure}
\begin{center}
 \includegraphics[width=8.5cm,height=3.5cm]{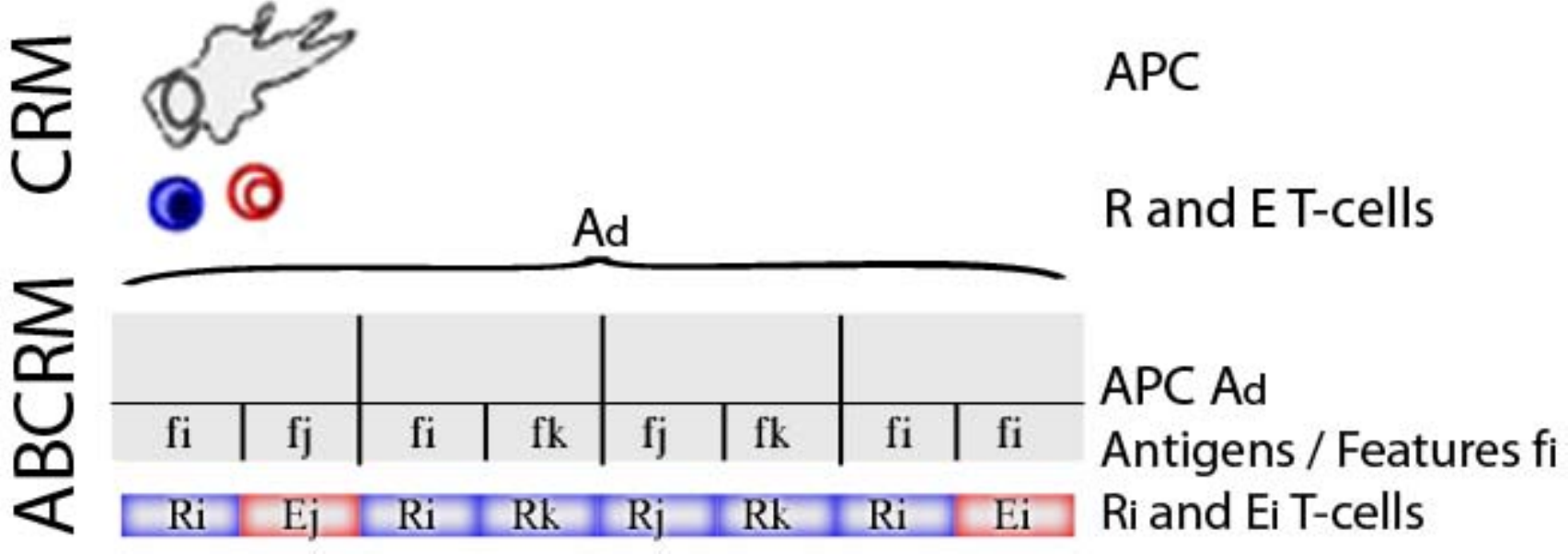}
\caption{ To illustrate the difference between the CRM and the ABCRM, the top
part of the figure represents a single APC of the CRM which can
bind to a maximum of two T-Cells. The lower part represents the APC for a document
$d$ in the ABCRM, which contains many pairs of antigen/feature ``slots'' where pairs
of T-cells can bind. In this example, the first pair of slots of the APC $A_d$ presents
the features $f_i$ and $f_j$; a regulatory T-cell $R_i$ and an effector
T-cell $E_j$ bind to these slots, which will therefore interact according to
reaction (4)---$R_i$ inhibits $E_j$ and in turn proliferates by doubling. The next pair
of slots leads to the interaction of regulatory T-cells $R_i$,$R_k$ that duplicate
via reaction (2)...}
\label{ada}
\end{center}
\end{figure}

In the natural immune system,  millions of novel T-cells are randomly generated in the thymus
every day to attempt to predict future antigens. In our algorithm, in contrast,
we generate T-cells only for features (words) occurring in the relevant
document corpus.
This is reasonable because the space of meaningful words in a language is
largely fixed and much smaller than the space of possible polypeptide epitopes
in biology.
More specifically, a document $d$ contains a set of features $F_d$; an artificial APC $A_d$ that
represents $d$, presents a subset of antigens/features  $A_d \subseteq F_d$  to artificial $E$
and $R$ T-cells. $E_f$ and $R_f$ bind to a specific feature $f$  on
\underline{any} APC that presents it; if  $f \in A_d$, then any available $E_f$ or $R_f$ in the cellular dynamics may bind stochastically to $A_d$\footnote{Every $E_f$ or $R_f$ has equal probability of binding to the APC that presents feature $f$}, as illustrated in figure \ref{ada}.

In biology, antigen recognition is a more complex process than mere polypeptide
sequence matching, but for simplicity we limit our feature recognition to
string matching.
APC are organized as a list of pairs of ``slots'' of (textual) features,
where T-cells, specific for those features, can bind. We use this
antigen/feature presentation scheme of pairs of ``slots'' to simplify our
algorithm. In future work we will study alternative feature presentation
scenarios.
An APC is modeled as a list of ``slots'' of pairs of features: $A_d = s_1 \cdots s_{n_S}$, where $s_d = \langle f, g \rangle$, $f, g \in A_d$, and $n_S = {{n_A \times |A_d|} \over {2}}$.
$f$ and $g$ are sampled (without repetition) from $A_d$ and randomly distributed exactly $n_A$ times over the list of slots that makes up the APC. Features are treated as \emph{bag of words}--i.e. the sequence of words in the document is not maintained \cite{feldman2006tmh}.
Once T-cells bind to an APC, every pair of
T-cells that binds to the same slot $s_d$ duplicates according to reaction rules (2-4).

In summary, each T-cell population is specific to and can bind to only one
feature presented by any APC. Implementing the algorithm as an Agent-based
model (ABM) allows us to deal with the recognition and co-recognition
(co-occurrence in the same document) of many features simultaneously,
rather than a single one as the original CRM does.

\begin{figure}[ht]
\hspace{-1cm}
\includegraphics[width=10.5cm, height=10.5cm]{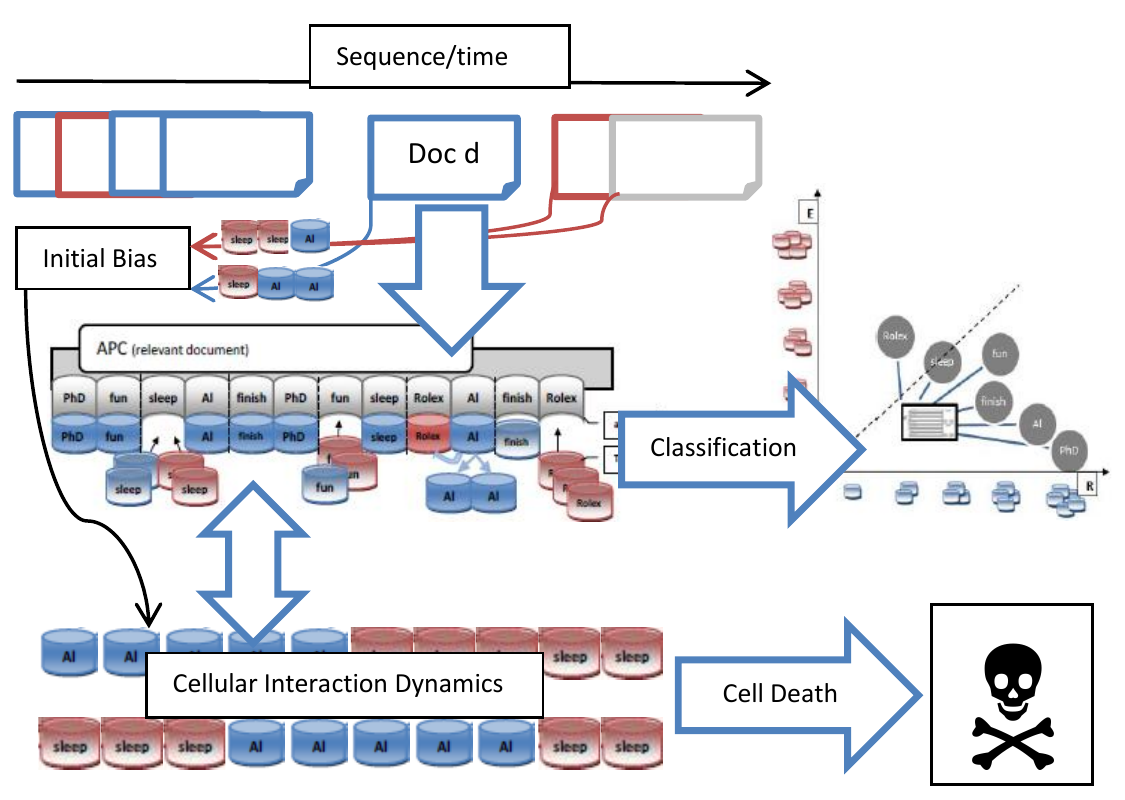}
\caption{ A stream of ordered labeled documents (blue for relevant and red for irrelevant) followed by ordered unlabeled grayed documents is introduced. Each document $d$ is represented by a polyspecific APC $A_d$ that arbitrarily presents the antigens/features $f$ of $d$. APC are then dropped in the pool of T-Cell populations representing previously encountered features/antigens, which follow the cellular interaction dynamics defined by the four interaction rules(see eq (2-4)). Finally, document $d$ is classified as relevant if the majority of its features $f$ have more $R_f$ than $E_f$, and irrelevant otherwise.}

\label{fola}

\end{figure}

The ABCRM uses incremental learning to first train on $N$ labeled documents
(relevant and irrelevant), which are ordered sequentially (typically by time
signature) and then test on $M$ unlabeled documents that follow in time order. Fig. \ref{fola} illustrates this stream of labeled documents (blue for relevant and red for irrelevant) followed by unlabeled  grayed documents.
The sequence in which documents are received affects the artificial cellular
dynamics, as incoming APC and T-cells face a T-cell dynamics that depends on
the specific documents previously encountered. Therefore, we use
publication-time as the default ordering for incoming documents, and study
if there is an advantage to preserving the original temporal sequence of
articles (see section \ref{OSO}).

Carneiro et al \cite{carneiro2007tnc} show that both $E$ and $R$ T-cells
co-exist in healthy individuals assuming enough APC exist. $R$ T-cells require
adequate amounts of $E$ T-cells to proliferate, but not too many that can
out-compete $R$ for the specific features presented by APC. ``Healthy'' T-cell
dynamics is identified by observing the co-existence of both $E$ and $R$
T-cells with $R \ge E$. ``Unhealthy'' T-cell dynamics is identified by
observing $E \gg R$, and should result when encountering many irrelevant
features in a document---in analogy with encountering many nonself antigens.

\begin{figure}[ht]
\begin{center}
\includegraphics[width=9cm]{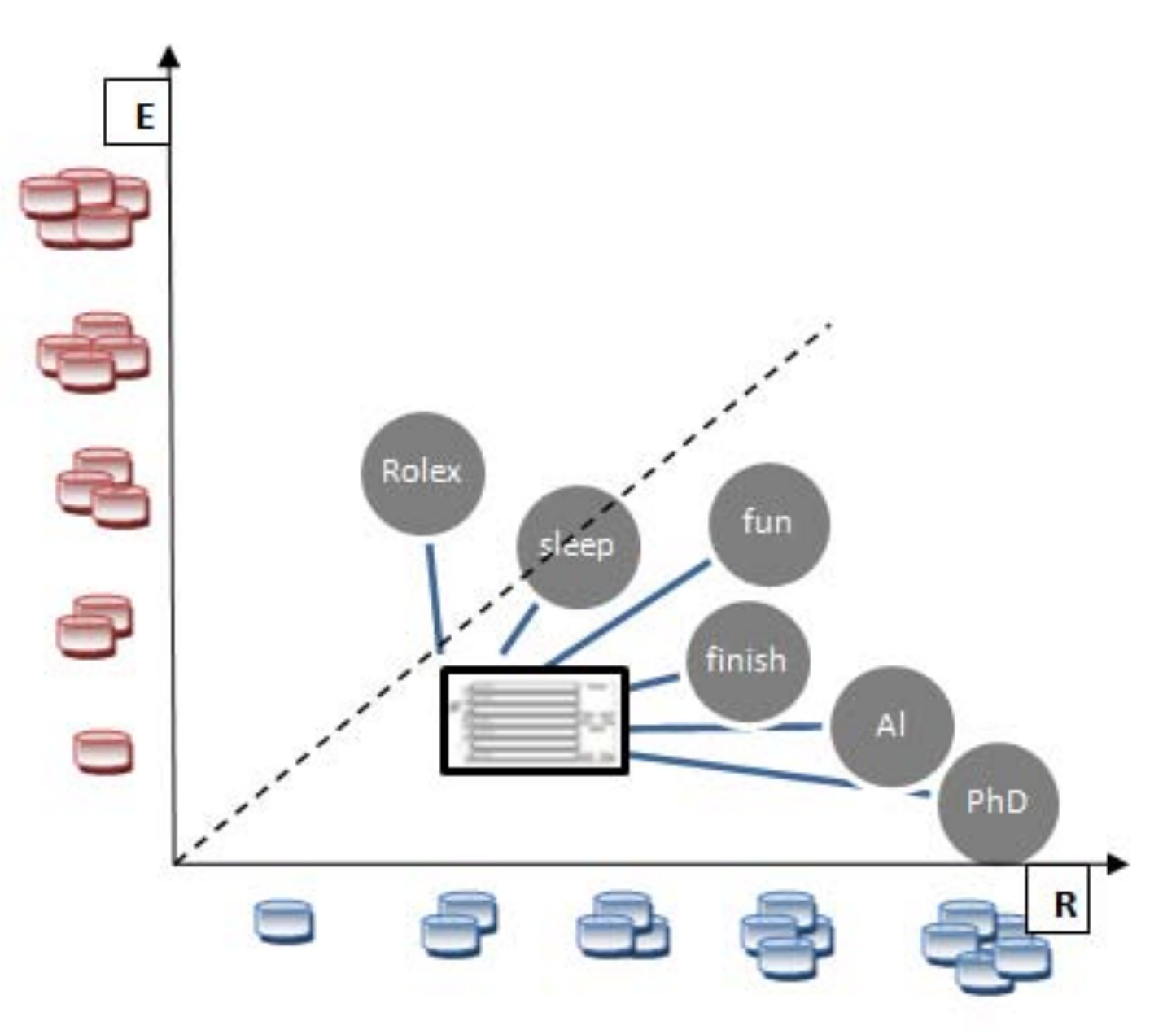}
\caption{ A document is classified according to the $E$-to-$R$ ratios for all its features. In this example, the e-mail document is classified as relevant given its features that tend to have higher ratios of $R$.}

\label{fclassification}
\end{center}
\end{figure}

In other words, features associated with relevant documents should have more
$R$ T-cell representatives than $E$ ones in the artificial cellular
dynamics. In contrast, features associated with
irrelevant documents should have many more $E$ than $R$ T-cells. Therefore,
when a document $d$ contains features $F_d$ that bind mostly to $E$ rather than
$R$ cells, we can classify it as irrelevant---and relevant in the opposite
situation (see Fig. \ref{fclassification}).

%
%

\vspace{0.2cm}

The ABCRM is  controlled by 6 parameters: \vspace{-0.1cm}
\renewcommand{\labelitemi}{$\bullet$}

\begin{itemize*}
\item \textbf{$E_0$} is the initial number of Effector T-cells generated for all new features
\item \textbf{$R_0^-$} is the initial number of Regulatory T-cells  generated for all new features in irrelevant and unlabeled (test) documents
\item \textbf{$R_0^+$} is the initial number of Regulatory T-cells  generated for all new features in relevant documents
\item \textbf{$d_E$} is the death rate for Effector T-cells that do not bind to APC
\item \textbf{$d_R$}  is the death rate for Regulatory T-cells that do not bind to APC
\item \textbf{$n_A$} is the number of total slots in which each feature $f$ is presented on APC
\end{itemize*}

When (textual) features are encountered for the first time, a fixed initial
number of $E_0$ effector T-Cells and $R_0$ regulatory T-Cells is generated for
every new feature $f$.
These initial values of T-cells vary for relevant and irrelevant documents in
training and in test stages. More Regulatory ($R_0^+$) than Effector T-cells
are generated for features that occur for the first time in documents that are
labeled relevant in the training stage ($R_0^+>E_0$), while fewer Regulatory
($R_0^-$) than Effector T-cells are generated in the case of irrelevant
documents ($R_0^- < E_0$) (see Fig. \ref{fola}). Features appearing in unlabeled documents for the
first time during the test stage are treated as features from irrelevant
documents, assuming that new features are irrelevant (nonself) until
neutralized by the collective dynamics given their co-occurrence with relevant
ones.

Naturally, relevant features will occur in irrelevant documents and vice versa.
However, the assumption is that relevant features tend to co-occur more
frequently with other relevant features in relevant documents and similarly for
irrelevant features.
Therefore, the proliferation dynamics defined by the 4 reactions and guided by
co-binding to APC slots is expected to correct the erroneous initial bias as we will show in section \ref{OIB}.
But this self-correction has not been proven in our previous works \cite{abihaidar_alife10,abihaidar_icaris10}, and it is one of the issues we
test in the present work.
%


Finally, to classify a document $d$, we observe the cellular interaction dynamics that results after its respective APC $A_d$ is left to interact with the various T-Cell populations. More specifically, each document is classified based on the $E$-to-$R$ ratios of all its features $f \in A_d$; this process is illustrated in Fig. \ref{fclassification}. A detailed pseudocode of the algorithm follows:


\newpage

 \textbf{ ABCRM Algorithm:}\\

\begin{algorithm}[H]
\SetLine
\KwIn{Stream of labeled and unlabeled documents}
\KwOut{Labels for unlabeled documents}

\ForEach {document $d$}{
Generate a list of pair slots $A_d$ presenting each $f \in A_d$ at $n_A$ randomly distributed slots.\\
Let $C$ contain $E_f$ and $R_f$ T-cells for all features $f$ in the cellular dynamics. \\

\ForEach {$f \in A_d$ representing document $d$}{
\If {$E_f \notin C$ and $R_f \notin C$}{
$E_f = E_0$  (i.e. generate $E_0$ Effector T-cells for $f$)

\If {$d$ is labeled relevant}{
$R_f = R_0^+$ (i.e. generate $R_0^+$ Regulatory T-cells for $f$)
}
\Else{
$R_f = R_0^-$ (i.e. generate $R_0^-$ Regulatory T-cells for $f$)
}
Update $C$ with $E_f$ and $R_f$ \\
Let all $E_f$, $R_f$ bind specifically to matching $f$ on $\mathbf{A_d}$: \\
}
}

\ForEach {pair of adjacent $(f, g)$ on $\mathbf{A_d}$}{
Apply the following interaction rules and update total number of $E$, $R$ T-cells: \\
     $(R_f,R_g) \rightarrow R_f + R_g$ \\
     $(E_f,E_g) \rightarrow 2.E_f + 2.E_g$\\
     $(E_f,R_g) \rightarrow E_f + 2.R_g$\\
}

\ForEach {$ R_f, E_f \in C$ that do not bind to $\mathbf{A_d}$} {
Cull $E_f$ and $R_f$ according to death rates $d_E$ and $d_R$\\
}
\If {$d$ is unlabeled}{
 Let $R(d) = \sum_{f \in A_d}({R_f \over \sqrt{R_f^2 + E_f^2}})$ and $E(d) = \sum_{f \in A_d}({E_f \over \sqrt{R_f^2 + E_f^2}})$\\
%
\If {$R(d) \ge E(d)$} {
Classify $d$ as relevant\\
}
\Else{
Classify $d$ as  irrelevant. \\
}
}
}

\end{algorithm}

\section{Data and Feature Selection}
\label{datas}
\begin{figure}
\begin{center}
\includegraphics[height=4cm,width=8cm]{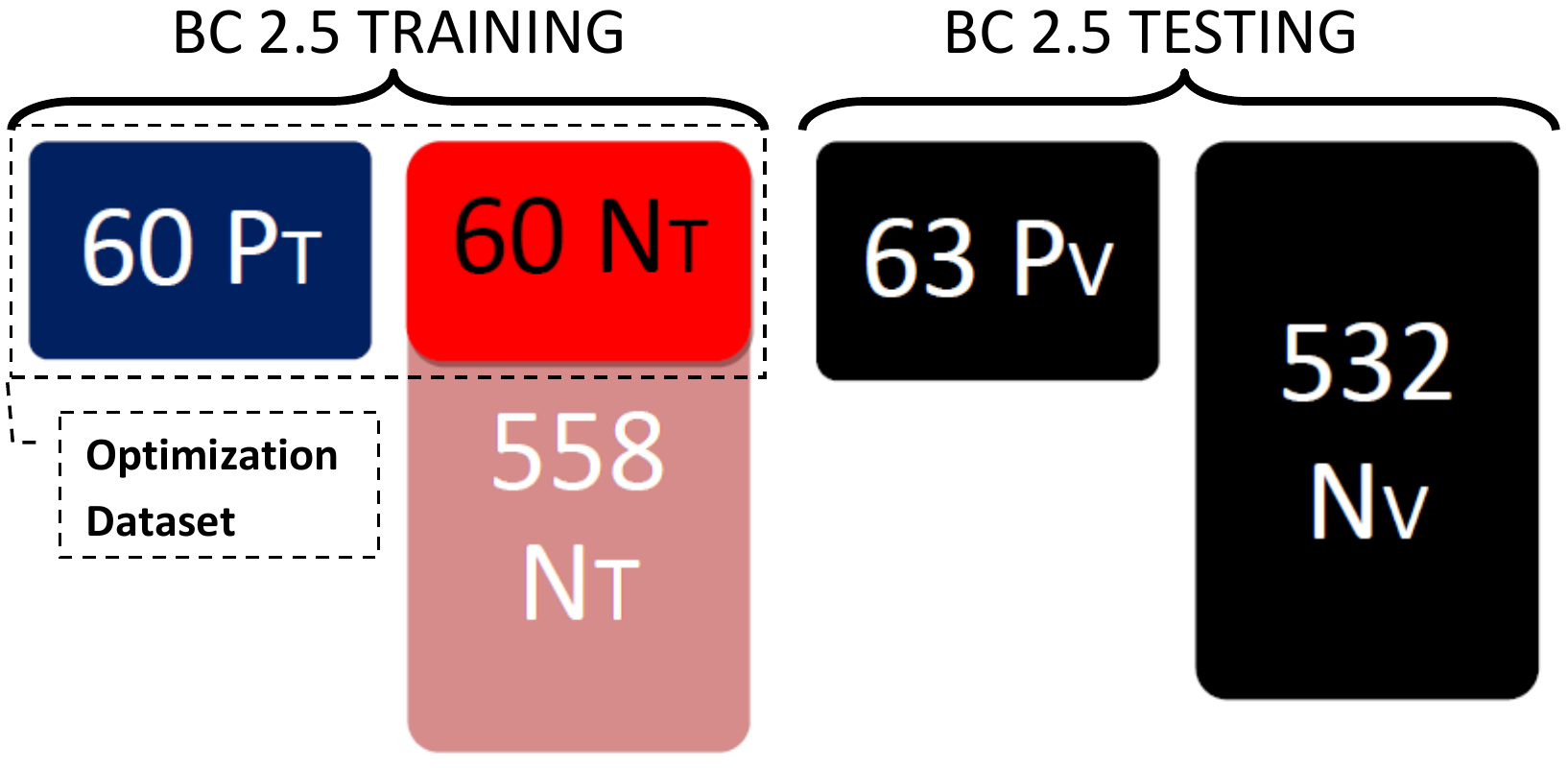}
\caption{ Numbers of relevant ($P$) and irrelevant ($N$) documents in the
training ($T$) and test ($V$) data sets of the Biocreative 2.5 challenge. In the optimization and robustness analysis stage, we use a balanced set of 60 $P_T$ (blue) and 60 $N_T$ (red) randomly selected
articles from the training data set and we call this subset the optimization dataset. In the test stage we use the unbalanced
validation set containing 63 $P_V$ (black) and 532 $N_V$ (black) documents. Notice that
the validation data was provided to the participants in the classification task of
Biocreative 2.5 unlabeled, therefore participants had no prior knowledge of
class proportions. }
\label{data}
\end{center}
\vspace{-0.5cm}
\end{figure}

The BioCreative (BC) challenge aims to assess the quality of
biomedical literature mining algorithms such as article classifiers.
The article classification task of Biocreative 2.5  \cite{bc2.5} was based on
a training data set ($T$) comprised of 61 full-text articles relevant ($P_T$)  to the
topic of \emph{protein-protein interaction} (PPI) and 558 irrelevant ones ($N_T$).
The realistic imbalance between the relevant and irrelevant instances is very
challenging for common machine learning techniques, since there are few
instances of the topical category of interest to generalize from.
Because we cannot predict how imbalanced the validation set will be, we first
search for optimal ABCRM parameters on a smaller sample of the training that is
balanced in the numbers of relevant and irrelevant documents.
The optimal parameters are not only useful for fine-tuning our algorithm for the best classification performance
but also for studying the robustness and behavior of T-cell dynamics under several experimental setups as we will show in section \ref{results}.
%
%
For this purpose, we chose the first 60 relevant and sampled 60 irrelevant
articles that were published around the same date (uniform distribution between
Jan and Dec 2008), and we called this subset the optimization dataset as illustrated in figure \ref{data}.
For final validation we used the entire Biocreative 2.5 test data set ($V$)
consisting of 63 full-text articles relevant to PPI ($P_V$) and 532 irrelevant
ones ($N_V$) as also shown in figure \ref{data}.
Furthermore, we compared our optimized algorithm with a Naive Bayes (NB)
\cite{metsis2006sfn} and a support vector machine (SVM) classifier
\cite{joachins_SVM_02}.


\begin{figure}[!ht]
\includegraphics[height=5cm, width=8.5cm]{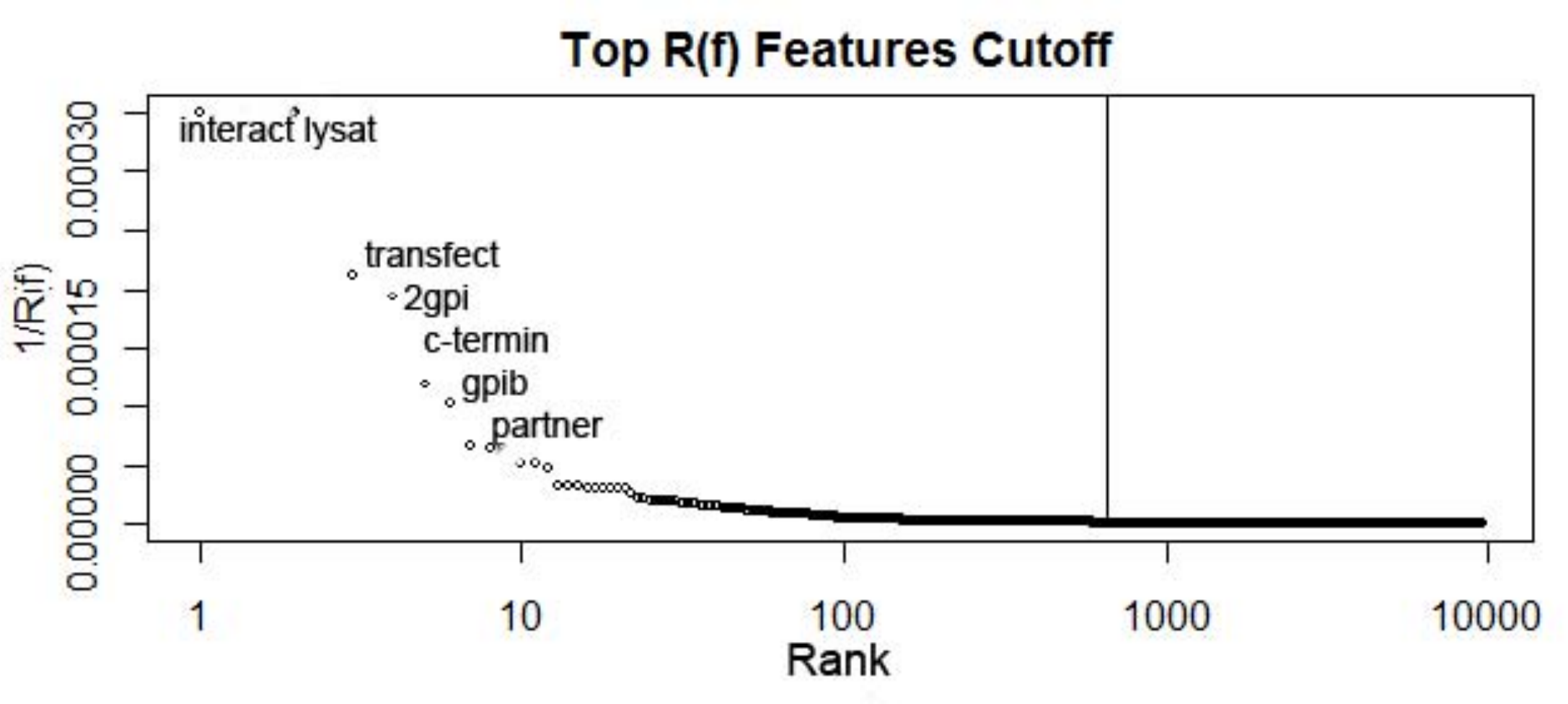}
\caption{We choose the top 650 ranked features according to the rank product R(f) = TF.IDF(f) $\times$ S(f). The y-axis represents $1 \over R(f)$ and the x-axis represents the index of R(f) for the sorted features. Features ranked below the 650th feature have a similar score ${1 \over R(f)} < 0.00001$ }
\label{cut}
\end{figure}

We pre-processed all articles by filtering out common words\footnote{The list
of common (stop) words includes 33 of the most common English words from which
we manually excluded the word ``with'', as we know it to be of importance to
PPI} and porter stemming \cite{porter1980} the remaining words which are all
the potential features.
We then ranked words/features $f$ extracted from training articles
($T$)\footnote{For feature extraction we used both the training data of
Biocreative 2.5 and Biocreative 2 as described in \cite{casci_TCBB_10}; all
classifiers used the exact same feature set.} according to two scores:
the first one is the average TF.IDF\footnote{TF.IDF is a common text weighting
measure to evaluate the importance of a feature/word in a document in a corpus.
TF stands for term frequency in a document and IDF for inverse document
frequency in the corpus.} \cite{feldman2006tmh}, and the
second one is the separation score  $S(f) = |p_{P}(f) - p_{N}(f)|$ where
$p_{P}$ $\left(p_{N}\right)$ is  the probability of a feature occurring in a
relevant (irrelevant) document of the training set $T$
\cite{abihaidar_GB08,casci_TCBB_10}. The two scoring and feature selection methods are useful for topical categorization but can be replaced by other methods to suit various applications that are beyond the focus of this manuscript.
The final rank $R(f)$ for every feature $f$ is given by the product of the
ranks obtained from both scores; we used only the top 650 ranked features
according to $R(f)$. These top 650 features were shown to be adequate for the classification of the same data set using a linear classifier \cite{casci_TCBB_10}. Moreover, a fixed number of features renders the algorithm more scalable for larger data sets with many more features, unlike the one used for this experiment. For example, features such as ``interact", ``lysat" and ``transfect" were ranked above others for their high ranks according to both scores as shown in figure \ref{cut}. See
\cite{casci_TCBB_10} for more details about the feature extraction procedure.

%

\noindent
\begin{table}
\begin{center}
\scalebox{1}{
 \begin{tabular}{| c | c | c | }
\hline
Parameter & Range & Step\\ \hline
$E_0$ &[1,7] & 1 \\
$R_0^-$ & [3,12] &1 \\
$R_0^+$ & [3,12] &1\\
$d_E$&[0.0,0.4] &0.1\\
$d_R$&[0.0,0.4] &0.1\\
$n_A$ & [2,22] & 2  \\ \hline
\end{tabular}}
\end{center}
\caption{ Parameter ranges used for optimizing the ABCRM  }
\label{pr} \vspace{-0.5cm}
\end{table}

\vspace{-0.5cm}

\section{Parameter Search and Robustness}
\label{results}

We performed an exhaustive parameter search by training the ABCRM on 60
balanced full-text articles (30 $P_T$ and 30 $N_T$ from BC2.5 training) and testing
it on the remaining 60 balanced ones (also 30 $P_T$ and 30 $N_T$ from BC2.5 Training)
as illustrated in figure \ref{data}\footnote{Notice that this parameter search
on the provided labeled training data uses only the information available to
the teams participating in Biocreative 2.5 challenge, and none of the test
data whose labels were revealed post-challenge.}.
Each run corresponds to a unique configuration of the 6 parameters of the
ABCRM. The explored parameter ranges are listed in table \ref{pr} and they result
in a total of 192500 unique parameter configurations for each experiment.
Finally, the parameter configurations were sorted with respect to the resulting
F-score measure of performance\footnote{F-score  = ${{2.Precision.Recall} \over
{Precision+Recall}}$ where Precision = ${{TP} \over {TP+FP}}$ and Recall =
${{TP} \over { TP+FN}}$. True Positives (TP) and False Positives (FP) are the
classifier's correct and incorrect predictions for relevant documents, while
True Negatives (TN) and False Negatives (FN) are the correct and incorrect
predictions for irrelevant documents.}, which is a good measure between
precision and recall when applied to balanced data \cite{beyond}.
%

We compiled the performance of the ABCRM on the entire parameter search space
for four distinct  experiments:  (1) the effect of \textbf{cell death}, (2)  using both \textbf{training sets} in contrast to using only the positive set, (3)  the importance of the \textbf{sequential order} of
articles, and (4) the automatic correction of the \textbf{initial bias}.

In all four experiments, we choose the 50 configurations with highest F-score
measure to study the ABCRM performance, because we are interested in
identifying the experimental setups that lead to higher \textbf{robustness} to
parameter changes.
We compare experimental outcomes with the paired student t-test; the null
hypothesis is that the two samples are drawn from the same distribution. A
p-value $< 0.01$ rejects the null hypothesis, establishing a statistical
distinction between the data drawn from two experimental setups---in our case,
the data from each experiment are the top 50 F-score values obtained.
The first two experiments were initially tested \cite{abihaidar_icaris10} to choose the best experimental set up and compare it with two aditional experiments \cite{abihaidar_alife10} that are discussed in this paper.

\subsection{Cell Death}
\label{OCD}

The \textbf{first} experiment aims to study the effect of cell death on immune memory and classification performance. In this experiment we compare the top 50 parameter configurations according to F-score obtained using cell death (exp 1.1) to those with no cell death (exp 1.2)---while training on both self and nonself documents. We observe a notable difference in classification performance that we validate statistically (according to the criteria above) to show that using cell death improves the performance (see Fig. \ref{fig_parameter_search_results})---regardless of whether the algorithm is trained on just relevant or on both relevant and irrelevant documents (see below). Therefore we conclude that cell death, which helps in the forgetting of useless features and focuses on more recent and frequent ones, improves classification performance, which suggests that it is important for immune memory in the T-Cell cross-regulation model.

\subsection{Training on Self and Nonself}
\label{OTS}

The \textbf{second} experiment is conducted to show if we can rely solely on the positive set for classification, or if the performance can be improved by training on both positive and negative sets.  We compare the top 50 parameter configurations according to F-score obtained using training on positive only or PU learning (experiments 2.1 and 2.2), to the previous experiments (1.1 and 1.2).
This way we compare training on positive documents only, with and without cell death.
The results show that using both training sets always (significantly) improves the robustness of classification performance  (see Fig. \ref{fig_parameter_search_results}).
Although the top performance obtained for 1.1 (training on both classes with cell death) and 2.1 (training on positive documents with ceall death) is equivalent with F-Score=0.85 (see table \ref{top_classifiers1}), the robustness as measured by the performance of the top 50 parameter sets is significantly lower for experiment 2.1 (see figure \ref{fig_parameter_search_results}).  
%

%
\noindent
\begin{table}
\begin{center}
\scalebox{1}{
 \begin{tabular}{| c || c | c | c | c | c | c | c |}
\hline
Exp. &  F-Score & $E_0$ & $R_0^+$ & $R_0^-$ & $d_R$ & $d_E$ & $n_A$ \\ \hline
$1.1$ & 0.85 & 2 & 11 & 10 & 0.3 & 0.2 & 18 \\
$1.2$ &  0.83  & 1 & 4 & 7 & 0.0 &  0.0  & 18  \\
$2.1$ & 0.85 & 1 &12 &8& 0.1& 0.0& 8   \\
$2.2$ & 0.75  &2 &12 &6 &0.0& 0.0 &18  \\

\hline
\end{tabular}
}
\end{center}
\caption{ Performance and parameters of top classifiers in
experiment 1 regarding cell death and experiment 2 regarding training data. } \label{top_classifiers1} \vspace{-0.5cm}
\end{table}

\noindent
\begin{figure}[!ht]

\noindent
\begin{center}
\includegraphics[width=9cm, height=6cm]{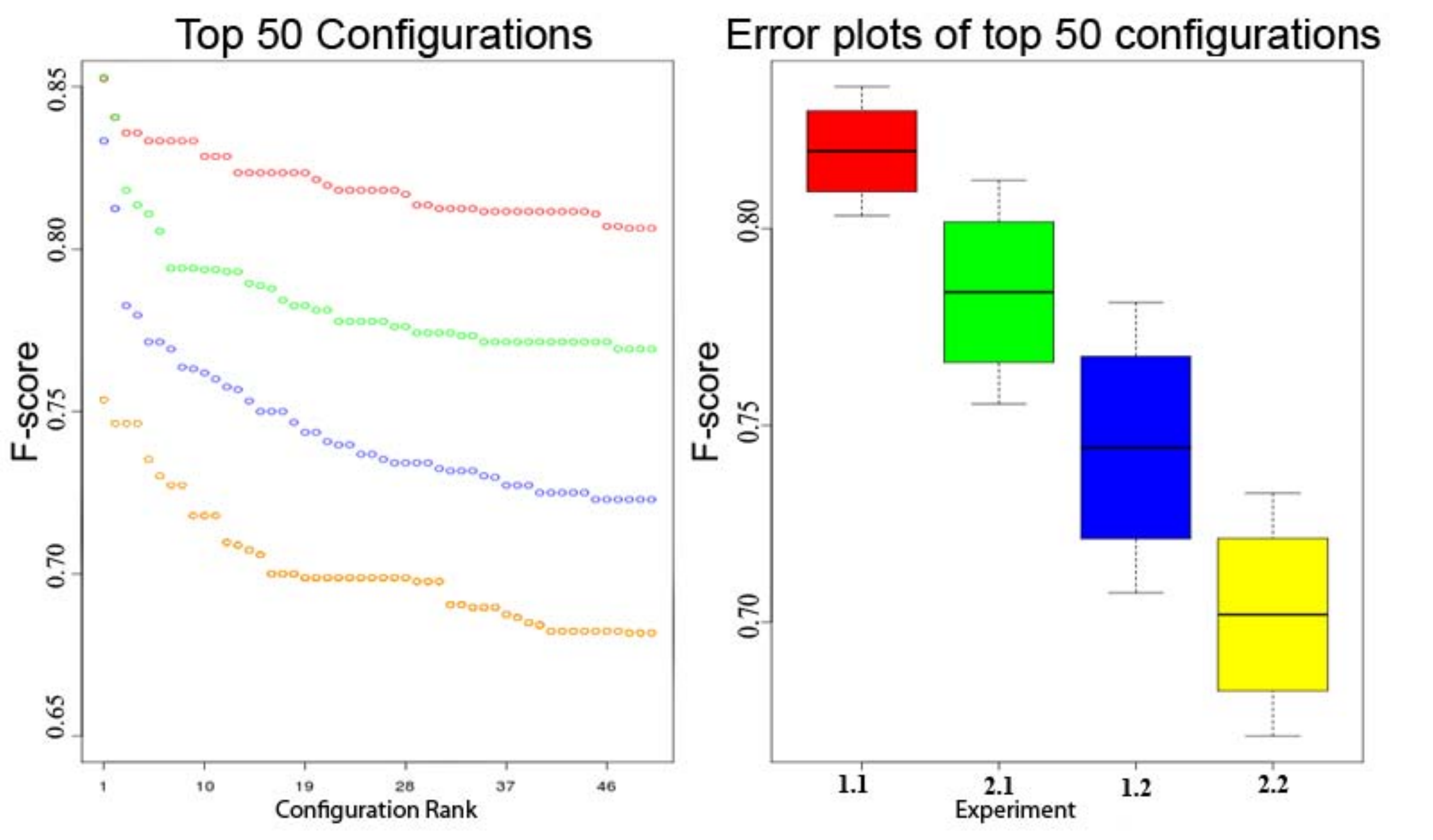}
\caption{The first two experiments result in four experimental setups: 1.1) training on both sets with cell death (red), 2.1) PU learning with cell death (green), 1.2) training on both sets with no cell death (blue) and 2.2) PU learning with no cell death (yellow) are clearly distinguishable for the top 50 configurations of each experiment on the plot on the left. On the right, the horizontal lines represent the mean, the boxes represent 95$\%$CI, and the whiskers represent standard deviation of F-scores from the top 50 parameter configurations}
\end{center}
\label{fig12}
\end{figure}

\subsection{Sequence Order}
\label{OSO}

The \textbf{third} experiment aims to establish how much the sequence order of
processing documents impacts performance. In particular, we test if preserving
the original temporal order of biomedical documents results in better
performance, as this would indicate that the ABCRM can use its
sequence-dependent dynamics to track the natural concept or topical drift and
thus improve classification.
Therefore, we compared the performance of the ABCRM when tested on a sequence
of biomedical articles ordered by the original publication, against randomly
shuffling the articles.
We tested four distinct experimental setups in order to fully explore the
influence of document order: \vspace{-0.1cm}

\begin{enumerate*}
\item  Ordered training set $\Rightarrow$ ordered test set
\item  Ordered training set $\Rightarrow$  shuffled test set
\item  Shuffled training set $\Rightarrow$  shuffled test set
\item  Shuffled training set $\Rightarrow$  ordered test set
\end{enumerate*}
\vspace{-0.1cm}

 In the case of shuffled sets, we produced 8 runs with distinct random document
orderings; in those cases, performance is represented by central tendency.
%

%


\begin{figure}[!ht]
\includegraphics[height=7cm,width=8.5cm]{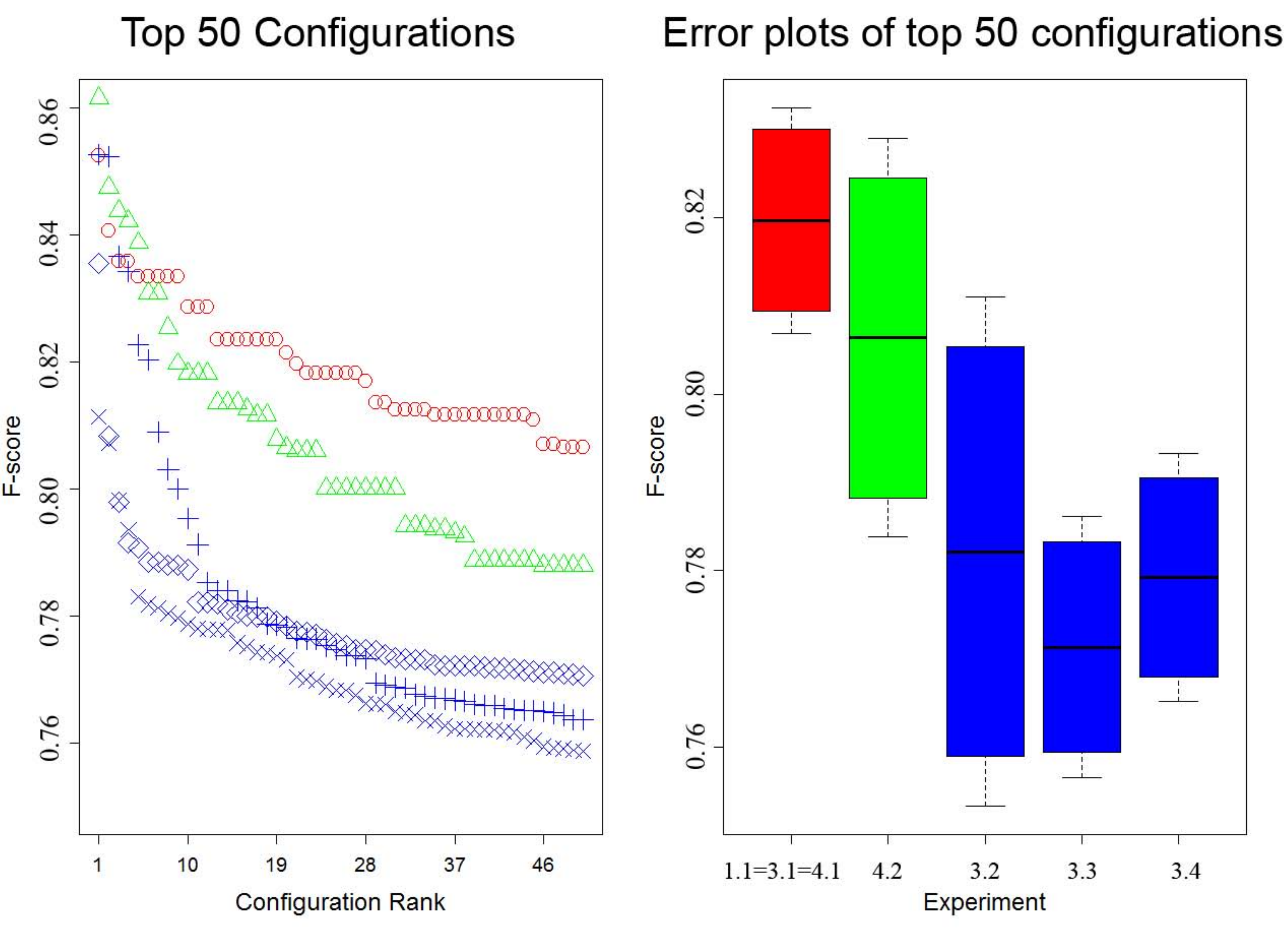}
\caption{ The second two experiments result in 5 experimental outcomes. To the left we show the top 50 parameter configurations ranked in terms of F-score for experimental setups 1.1=3.1=4.1 (red circles), 3.2 (blue pluses), 3.3 (blue crosses), 3.4 (blue diamonds), and 4.2 (green triangles). To the right we show the mean (line), 95$\%$CI (boxes), and standard deviation (whiskers) of F-scores for the top 50 parameter configurations.}
\label{fig_parameter_search_results}
\end{figure}

\noindent
\begin{table}
\begin{center}
\scalebox{1}{
 \begin{tabular}{| c || c | c | c | c | c | c | c |}
\hline
Exp. &  F-Score & $E_0$ & $R_0^+$ & $R_0^-$ & $d_R$ & $d_E$ & $n_A$ \\ \hline
$1.1=3.1=4.1$ & 0.85 & 2 & 11 & 10 & 0.3 & 0.2 & 18 \\
$3.2$ & 0.85 & 2 & 7 & 6 & 0.0 & 0.0 & 20 \\
$4.2$ & 0.86 & 3 & 8 & 7 & 0.2 & 0.1 & 14 \\ \hline
\end{tabular}
}
\end{center}
\caption{ Performance and parameters of top classifiers in
experiments 1.1=3.1=4.1, 3.2 and 4.2. } \label{top_classifiers} \vspace{-0.5cm}
\end{table}

\vspace{-0.5cm}

The results of  this experiment are summarized in figure \ref{fig_parameter_search_results}.
The robustness of performance of the first experimental setup (preserving
temporal order of articles) is significantly above the other setups.
Using the paired student t-test as described above,
we conclude that the ABCRM is sensitive to article order---i.e. if the articles
are shuffled, the performance is worse.
While the performance of the best classifier obtained via experimental setup
3.2 is equivalent to the best one obtained for experimental setup 1.1 (F-Score
= 0.85, see table \ref{top_classifiers} and figure
\ref{fig_parameter_search_results}), that setup is very sensitive to parameter
changes and the performance quickly and significantly decreases for subsequent
best classifiers (see figure \ref{fig_parameter_search_results}). Indeed, the
performance of the top 50 classifiers for experimental setups 3.2, 3.3, and 3.4
is statistically indistinguishable from each other, but is significantly lower
than the performance of the top 50 classifiers for experimental setup 1.1.
This means that there is indeed a conceptual drift in the Biocreative 2.5
article data stream, and the ABCRM can track it better (and in a more robust
manner) when publication date is used as the sequence for processing articles
than when the temporal order of articles is shuffled.
This also suggests that the process of T-Cell cross-regulation in the IS, as
modeled here, can track changing nonself pathogens. 
%
%

It should be noted that in this experiment, the partitioning of training and test data was done according to the time-stamp of documents. Therefore, the documents in the test set were published after all documents in the training set.
Therefore, even in the shuffled training and test sets (experimental setup 3.3), there is some preservation of temporal order.
In future work we will explore experimental setups where the training and test sets are drawn from the same time-stamp distribution to better understand the effects of concept drift and how well our model can track it.
%


\subsection{Initial Bias}
\label{OIB}

In the \textbf{fourth} experiment we test the effect of the initial biases
introduced when features are first encountered. The initial biases of
regulatory T-cells injected in the dynamics for a new feature $f$, depend on
whether the first document $d$ where the feature is encountered is labeled
irrelevant/unknown ($R_0^-$) or relevant ($R_0^+$).
Since features will occur in both relevant and irrelevant articles, this
initial bias for a feature could be detrimental, as a feature most associated
with one class could be first encountered on a document of the opposite class.
Therefore, it is important to test if the dynamics of the four reactions and
APC feature co-presentation that define the ABCRM can self-correct such
erroneous biases.
To perform this test, we altered the ABCRM algorithm such that T-cells are
incremented appropriately every time a feature occurs in a document, and not
just the first time the feature occurs (as the canonical algorithm does).
Specifically, every time a feature $f$ occurs in a document $d$, we increment
$E_f=E_f+E_0$ and $R_f=R_f+R_0^+$ if $d$ is labeled relevant and
$R_f=R_f+R_0^-$ if $d$ is labeled irrelevant or unlabeled.
We label this experimental set up 4.2, which was conducted with cell death and training on both positive and negative documents.


%
The results of this experiment are also summarized in figure
\ref{fig_parameter_search_results}.
The performance of top classifiers obtained for experimental setups 4.1 (same
as 1.1 and 3.1 that are trained on both training sets using cell death) and 4.2 (incremental experimental setup) is shown in table \ref{top_classifiers}. While the best overall
classifier is obtained with experimental setup 4.2, the performance of both
setups is statistically indistinguishable.
Indeed, using the paired student t-test as described above,
we cannot reject the null hypothesis claiming that both distributions of F-scores were drawn from a similar distribution. Therefore, we conclude that this modification does not improve the performance of the
ABCRM on the Biocreative data set, thus showing that the initial bias can be
corrected by the ABCRM collective dynamics and does not require incrementing T-cells for all new features.
Because features most associated with a given class tend to co-occur in text
with other features most associated with the same class, they will also tend to
be co-presented in APC and thus the relevant T-cells will proliferate with
similar rates. Therefore, the dynamics of the ABCRM can self-correct initial
erroneous biases from the natural textual co-occurrence of features.
This shows that T-Cell cross-regulation as modeled here can self-correct
initial antigen misclassification by the IS, assuming that antigens from one
class (self/nonself) tend to co-occur with antigens from the same class.

\section{Validation and Conclusions}
\label{valid}

To test the ABCRM on the full, unbalanced test set of the Biocreative
challenge (see  figure \ref{data}), thus establishing its merit as a
bio-inspired biomedical literature mining classifier, we adopted the best
parameter configuration from the canonical ABCRM (experimental setup 1.1=3.1=4.1, see table \ref{top_classifiers}) obtained from the parameter search
described above.
%
%
We compared the ABCRM classifier with the multinomial Naive Bayes (NB) with
boolean attributes, one of the top Naive Bayes implementations for spam detection \cite{metsis2006sfn}, and the publicly available
SVM$^{light}$ implementation of SVM applied to normalized feature counts
\cite{joachins_SVM_02}. The SVM$^{light}$ was used with its default parameter settings \cite{joachins_SVM_02}. All classifiers were tested on the same features
obtained from the same data.

\begin{table}[!ht]
\begin{center}
\scalebox{1}{
  \begin{tabular}{| c || c | c | c || c | c | c |}
    \hline
   & ABCRM & NB  & SVM  & Mean & StDev. & Med. \\ \hline
Precision & 0.22 &   0.14 &  0.24 & 0.38 & & \\
Recall & 0.65 &  0.71 & 0.94 & 0.68 & & \\ \hline
\textbf{F-score} & 0.33 & 0.24 & 0.36 & 0.39 & 0.14 & 0.38 \\
\textbf{Accuracy} & 0.71 &0.52 & 0.74 & 0.67 & 0.30& 0.84 \\
\textbf{AUC} & 0.34 & 0.19 & 0.46 & 0.43  & 0.17 & 0.44 \\
\textbf{MCC}&0.24&0.13&0.31& 0.31 & 0.19 & 0.33 \\
 \hline
  \end{tabular}
}
\end{center}

\caption{ F-Score, Accuracy, AUC and MCC performance of various
classifiers when training on the balanced training set of articles and testing
on the full unbalanced Biocreative 2.5 test set. Also shown is the central
tendency and variation of all systems submitted to Biocreative 2.5.}
\label{ress}
\end{table}

Since the F-score and Accuracy are not very reliable for evaluating unbalanced
classification \cite{beyond}, we also use the Area Under the interpolated
precision and recall Curve (AUC) and Matthew's Correlation Coefficient (MCC).
The results are listed in table \ref{ress}, which also includes the central
tendency of the results of all systems submitted by all Biocreative 2.5
participating teams \cite{bc2.5,casci_TCBB_10}.
It should be noted that the ABCRM, NB, and SVM classifiers we tested here, used
only single-word features because we wish to establish the feasibility of the
method. In contrast, most classifiers submitted to the Biocreative 2.5
challenge (including another method from our group which was one of the
top-performing classifiers \cite{casci_TCBB_10}) used more sophisticated
features such as bigrams and problem-specific entities. Therefore, it is not
surprising that these methods as tested here performed under the mean of the
challenge.
Our goal was to establish the ABCRM as a new bio-inspired text classifier to be
further improved in the future with more sophisticated features. When we
compare its performance to NB and SVM on the \underline{exact same} single-word
features, the results are encouraging. Indeed, based on the given measures,
while the SVM out-performed the ABCRM, the latter out-performed NB. 
%
%
Therefore, the
dynamics of T-Cell cross-regulation lead to a competitive collective
classification of biomedical articles, which we intend to develop further.

In future work we will pursue additional experiments to study concept drift, namely by investigating the ability to simultaneously train and classify documents. Given the sequence-dependent dynamics entailed by our model, there is no reason to present all test data to the cellular interaction dynamics, only after processing all training data. The model affords various possible schedules of document processing that mix training and test data which could lead to better performance.
Indeed, the immune system is constantly exposed to self antigens (training data), and even pathogens that may be stored in long lived plasma cells and memory B-cells.
%
%
%

In conclusion, we observed that our method uses cell death to enhance immune memory and forget older features while focusing on more recent and frequent ones. We proved that our algorithm is capable of classification when trained on relevant features only, however the performance can be improved when trained on both classes.
We also observed that algorithm adapts to the initial bias of
T-cell populations generated for new features,  and it performs best when
tested on a sequence of articles ordered by publication date---showing that it
can track concept drift in the biomedical literature.
%

These properties of our model also show that T-Cell cross
regulation is capable of efficient collective classification of nonself
antigens and suggest that T-Cell cross-regulation can naturally respond to
drift in the pathogen population.
Therefore T-Cell cross-regulation defined by the 4 reaction rules and
co-presentation of features in APC can be seen as an effective general
principle of collective classification available to populations of cells.
Clearly, there is still much to do to improve the model. For biomedical
literature mining applications, we need to test it with more sophisticated
features (as top classifiers in the field do). For our goal of understanding
T-Cell cross-regulation in the IS, we need to understand better how memory is
sustained in the collective cellular dynamics; for instance, how to sustain
regulatory T-Cells, which keep memory of self, in the dynamics even in the
presence of very unbalanced scenarios where there are many more self or nonself
instances. 

\begin{acknowledgements}
This work was partially supported by a grant from the FLAD Computational Biology Collaboratorium at the Instituto Gulbenkian de Ciencia in Portugal. We also thank the ICARIS2010 committee board for encouraging this work. We acknowledge the computational resources provided by Indiana University used to conduct the simulations we report.
\end{acknowledgements}

\bibliographystyle{spmpsci}      
\bibliography{EI}   

\begin{thebibliography}{}
\providecommand{\url}[1]{{#1}}
\providecommand{\urlprefix}{URL }
\expandafter\ifx\csname urlstyle\endcsname\relax
  \providecommand{\doi}[1]{DOI~\discretionary{}{}{}#1}\else
  \providecommand{\doi}{DOI~\discretionary{}{}{}\begingroup
  \urlstyle{rm}\Url}\fi

\bibitem{burnet1959clonal}
Burnet, S.F.M.: {The clonal selection theory of acquired immunity}.
\newblock Vanderbilt University Press (1959)

\bibitem{porter1980}
Porter, MF: An algorithm for suffix stripping.
\newblock Program \textbf{13}(3), 130--137 (1980)

\bibitem{paul1993fi}
Paul, W.E. and Technologies, I.O.: {Fundamental immunology}.
\newblock Raven Press New York (1993)

\bibitem{Crutchfield&Mitchell95}
James Crutchfield and Melanie Mitchell: The evolution of emergent computation.
\newblock PNAS \textbf{92}(23) (1995)

\bibitem{hofmeyr2001iii}
S.A. Hofmeyr: {An Interpretative Introduction to the Immune System}.
\newblock Design Principles for the Immune System and Other Distributed
  Autonomous Systems  (2001)

\bibitem{Segel&Cohen01}
Segel, L.A. and Cohen, I.: Design Principles for the Immune System and Other
  Distributed Autonomous Systems.
\newblock Oxford University Press (2001)

\bibitem{twycross2002isa}
Twycross, J. and Cayzer, S.: {An immune system approach to document
  classification}.
\newblock Master's thesis, COGS, University of Sussex, UK  (2002)

\bibitem{decastro}
De Castro, L.N. and Timmis, J.: {Artificial immune systems: a new computational
  intelligence approach}.
\newblock Springer Verlag (2002)

\bibitem{joachins_SVM_02}
T. Joachims: Learning to classify text using support vector machines: methods,
  theory, and algorithms.
\newblock Kluwer Academic Publishers (2002)

\bibitem{garrett2003pne}
Garrett, SM: {A paratope is not an epitope: Implications for immune networks
  and clonal selection}.
\newblock pp., 217--228 (2003)

\bibitem{Shatkay03JCB}
Hagit Shatkay and Ronen Feldman: Mining the biomedical literature in the
  genomic era: An overview.
\newblock Journal of Computational Biology \textbf{10}(6), 821--856 (2003)

\bibitem{bibliome_hersh04}
Hersh, William and Bhupatiraju, Ravi Teja and Corley, Sarah: Enhancing access
  to the bibliome: the trec genomics track.
\newblock Medinfo \textbf{11}(Pt 2), 773--777 (2004)

\bibitem{Peak:2004fk}
David Peak and Jevin D. West and Susanna M. Messinger and Keith A. Mott:
  Evidence for complex, collective dynamics and distributed emergent
  computation in plants.
\newblock PNAS \textbf{101}(4), 918--922 (2004)

\bibitem{Tsymbal2004}
Tsymbal, Alexey: The problem of concept drift: definitions and related work.
\newblock Computer Science Department Trinity College Dublin \textbf{4}(C),
  2004–15 (2004)

\bibitem{Rocha2005a}
Rocha, L.M. and Hordijk, W.: Material representations: From the genetic code to
  the evolution of cellular automata.
\newblock Artificial Life \textbf{11}(1-2), 189--214 (2005)

\bibitem{biocreative_issue}
Hirschman, Lynette and Yeh, Alexander and Blaschke, Christian and Valencia,
  Alfonso: Overview of biocreative: critical assessment of information
  extraction for biology.
\newblock BMC Bioinformatics \textbf{6 Suppl 1}, S1 (2005)

\bibitem{quorum_sensing_ant_Pratt2005}
Pratt, Stephen C.: Quorum sensing by encounter rates in the ant temnothorax
  albipennis.
\newblock Behav. Ecol. \textbf{16}(2), 488--496 (2005).
\newblock \doi{10.1093/beheco/ari0210.1093/beheco/ari020}

\bibitem{Shalizi:2006fk}
Cosma Shalizi and Rob Haslinger and Jean-Baptiste Rouquier and Kristina
  Klinkner and Cristopher Moore: Automatic filters for the detection of
  coherent structure in spatiotemporal systems.
\newblock Phys.Rev.E \textbf{73} (2006)

\bibitem{beyond}
Sokolova, M. and Japkowicz, N. and Szpakowicz, S.: {Beyond accuracy, f-score
  and roc: a family of discriminant measures for performance evaluation}.
\newblock pp.1015--1021 (2006)

\bibitem{hunter2006blp}
Hunter, L. and Cohen, K.B.: Biomedical language processing: What's beyond
  pubmed?
\newblock Molecular Cell \textbf{21}(5), 589--594 (2006)

\bibitem{Mitchell:2006fk}
Melanie Mitchell: Complex systems: Network thinking.
\newblock Artificial Intelligence \textbf{170}(18), 1194--1212 (2006)

\bibitem{Jensen2006}
Jensen, L. and Saric, J. and Bork, P.: Literature mining for the biologist:
  from information retrieval to biological discovery.
\newblock Nat Rev Genet \textbf{7}(2), 119--129 (2006).
\newblock \doi{10.1038/nrg1768}

\bibitem{Quorum_Sensing_Walters2006}
Matthew Walters and Vanessa Sperandio: Quorum sensing in escherichia coli and
  salmonella.
\newblock Int. Journal of Medical Microbiology \textbf{296}(2-3), 125 -- 131
  (2006).
\newblock \doi{DOI: 10.1016/j.ijmm.2006.01.041}

\bibitem{metsis2006sfn}
Metsis, V. and Androutsopoulos, I. and Paliouras, G.: {Spam Filtering with
  Naive Bayes--Which Naive Bayes?}
\newblock Third Conf. on Email and Anti-Spam (CEAS)  (2006)

\bibitem{feldman2006tmh}
Feldman, R. and Sanger, J.: {The Text Mining Handbook: advanced approaches in
  analyzing unstructured data}.
\newblock Cambridge University Press (2006)

\bibitem{timmis2007ais}
Timmis, J.: {Artificial immune systems today and tomorrow}.
\newblock Natural Computing \textbf{6}(1), 1--18 (2007)

\bibitem{Krallinger_BC2_IAS}
Martin Krallinger and Alfonso Valencia: Evaluating the detection and ranking of
  protein interaction relevant articles: the biocreative challenge interaction
  article sub-task (ias).
\newblock In: Proc. 2nd Biocreative Challenge Evaluation Workshop (2007)

\bibitem{carneiro2007tnc}
J. Carneiro and K. Leon and I. Caramalho and C. van den Dool and R. Gardner and
  V. Oliveira and M.L. Bergman and N. Sep{\'u}lveda and T. Paix{\~a}o and J.
  Faro and J. Demengeot: When three is not a crowd: a crossregulation model of
  the dynamics and repertoire selection of regulatory cd4 t cells.
\newblock Immunological Reviews \textbf{216}(1), 48--68 (2007)

\bibitem{abihaidar_icaris_08}
Alaa Abi-Haidar and Luis M. Rocha: Artificial Immune Systems (Proc. ICARIS).
\newblock pp., 36--47 (2008)

\bibitem{abihaidar_alifexi_08}
Alaa Abi-Haidar and Luis M. Rocha: Artificial Life XI: 11th Int. Conf. on the
  Simulation and Synthesis of Living Systems.
\newblock pp., 1--9. MIT Press (2008)

\bibitem{Helikar:2008fk}
Tom{\'a}s Helikar and John Konvalina and Jack Heidel and Jim A Rogers: Emergent
  decision-making in biological signal transduction networks.
\newblock Proc Natl Acad Sci U S A \textbf{105}(6), 1913--1918 (2008).
\newblock \doi{10.1073/pnas.0705088105}

\bibitem{dasgupta2008ict}
Dasgupta, D. and Nino, F.: {Immunological Computation: Theory and
  Applications}.
\newblock AUERBACH (2008)

\bibitem{abihaidar_GB08}
Alaa Abi-Haidar and Jasleen Kaur and Ana Maguitman and Predrag Radivojac and
  Andreas Retchsteiner and Karin Verspoor and Zhiping Wang and Luis M. Rocha:
  Uncovering protein interaction in abstracts and text using a novel linear
  model and word proximity networks.
\newblock p.9(Suppl 2):S11 (2008)

\bibitem{bc2.5}
Krallinger, M: The biocreative ii. 5 challenge overview.
\newblock p., 19 (2009)

\bibitem{nuno}
Nuno H. Sepulveda: How is the t-cell repertoire shaped.
\newblock Ph.D. thesis, Instituto Gulbenkian de Ciencia (2009)

\bibitem{abihaidar_icaris10}
Alaa Abi-Haidar and Luis M. Rocha: ICARIS 2010: Proc. of the 9th Int. Conf. on
  Artificial Immune Systems.
\newblock In: , pp., 237--249 (2010)

\bibitem{casci_TCBB_10}
Kolchinsky, Artemy and Abi-Haidar, Alaa and Kaur, Jasleen and Hamed, Ahmed
  Abdeen and Rocha, Luis M: {Classification of protein-protein interaction
  full-text documents using text and citation network features.}
\newblock IEEE/ACM transactions on computational biology and bioinformatics /
  IEEE, ACM \textbf{7}(3), 400--11 (2010).
\newblock \doi{10.1109/TCBB.2010.55}.
\newblock
  \urlprefix\url{http://www.computer.org/portal/web/csdl/doi/10.1109/TCBB.2010%
.55}

\bibitem{abihaidar_alife10}
Alaa Abi-Haidar and Luis M. Rocha: Artificial Life XII: Twelfth International
  Conference on the Simulation and Synthesis of Living Systems.
\newblock In: , pp., 706--713 (2010)

\end{thebibliography}

%
%

\end{document}